\documentclass[pre,superscriptaddress,twocolumn,letterpaper,nofootinbib]{revtex4}

\usepackage{graphicx}
\usepackage{amssymb}
\usepackage{chemarr}
\usepackage{appendix}
\usepackage{setspace}
\usepackage{color}
\usepackage{soul}
\usepackage{subfigure}

\DeclareMathOperator{\U}{\boldsymbol{U}}

\DeclareMathOperator{\m}{{\bf m}}
\DeclareMathOperator{\Uro}{\boldsymbol{U}^{\hbox{\tiny{(read)}}}}
\DeclareMathOperator{\UF}{\boldsymbol{U}^{\hbox{\tiny{(bSSFP)}}}}

\DeclareMathOperator{\Ufr}{\boldsymbol{U}^{\hbox{\tiny{(free)}}}}

\DeclareMathOperator{\TsM}{\mathit{T}_1^{*\hbox{\tiny{(MOLLI)}}}}
\DeclareMathOperator{\Tsb}{\mathit{T}_1^{*\hbox{\tiny{(bSSFP)}}}}
\DeclareMathOperator{\Tzsb}{\mathit{T}_2^{*\hbox{\tiny{(bSSFP)}}}}
\DeclareMathOperator{\TsLL}{\mathit{T}_1^{*\hbox{\tiny{(LL)}}}}
\DeclareMathOperator{\TsNCLL}{\mathit{T}_1^{*\hbox{\tiny{(NCLL)}}}}
\DeclareMathOperator{\tread}{\mathit{t}_{\hbox{\tiny{read}}}}
\DeclareMathOperator{\tfree}{\mathit{t}_{\hbox{\tiny{free}}}}
\DeclareMathOperator{\tim}{\mathit{t}_{\hbox{\tiny{im}}}}
\DeclareMathOperator{\TR}{\mathit{T}_{\hbox{\tiny{R}}}}
\DeclareMathOperator{\TE}{\mathit{T}_{\hbox{\tiny{E}}}}
\DeclareMathOperator{\TRR}{\mathit{T}_{\hbox{\tiny{RR}}}}
\DeclareMathOperator{\ez}{{{\bf e}}_z}
\DeclareMathOperator{\e3}{{{\bf e}}_\mathrm{3}}

\DeclareMathOperator{\mF}{{{\bf m}}_\mathrm{ss}^{\hbox{\tiny{(bSSFP)}}}}
\DeclareMathOperator{\mro}{{{\bf m}}_\mathrm{ss}^{\hbox{\tiny{(read)}}}}

\begin{document}

\title{An Analytical Model which Determines the Apparent $T_1$ for Modified Look-Locker Inversion Recovery  -\\ Analysis of the Longitudinal Relaxation under the Influence of Discontinuous Balanced (Classical MOLLI) and Spoiled Gradient Echo Readouts -\\ }

\author{Thomas Kampf}
\affiliation{
Institute of Experimental Physics V. University of W\"uzburg, Am Hubland, 97074 W\"urzburg}

\author{Theresa Reiter}

\author{Wolfgang Rudolf Bauer\footnote{corresponding author,
Tel. +49-931-201-39011}}
\email{bauer_w@ukw.de}
\affiliation{
Department of Internal Medicine I, University Hospital of W\"urzburg,
Oberd\"urrbacher Stra{\ss}e 6,
D-97080 W\"urzburg, Germany,\\
Comprehensive Heart Failure Centre, Am Schwarzenberg 15, r
D-97080 W\"urzburg, Germany
}

\date{\today}

\begin{abstract}
Quantitative nuclear magnetic resonance imaging (MRI) shifts more and more into the focus of clinical research. Especially determination of relaxation times without/and with contrast agents becomes the foundation of tissue characterization, e.g. in cardiac MRI for myocardial fibrosis.  Techniques which assess longitudinal relaxation times rely on repetitive application of readout modules, which are interrupted by free relaxation periods, e.g. the Modified Look-Locker Inversion Recovery = MOLLI sequence. These discontinuous sequences reveal an apparent relaxation time, and, by techniques extrapolated from continuous readout sequences, a putative real $T_1$ is determined. What is missing is a rigorous analysis of the dependence of the apparent relaxation time on its real partner, readout sequence parameters and biological parameters as heart rate. This is provided in this paper for the discontinuous balanced steady state free precession (bSSFP) and spoiled gradient echo readouts. It turns out that the apparent longitudinal relaxation rate is the time average of the relaxation rates during the readout module, and free relaxation period. Knowing the heart rate our results vice versa allow to determine the real $T_1$ from its measured apparent partner.

\end{abstract}
\maketitle

\section{Introduction}
Many nuclear magnetic resonance imaging techniques depend on periodic perturbative readouts of nuclear magnetization, the dynamics of which otherwise would be solely determined by thermodynamic forces driving it towards equilibrium. Prominent examples of periodic perturbations of relaxation processes are the repetitive application of - spoiled gradient echo sequences in order to determine quickly $T_1$ (Snapshot Flash) \cite{Deichmann1992} -- or balanced steady state free precession sequences (bSSFP)  \cite{Scheffler2003}. Recently characterization of myocardial pathology and function by fast determination of $T_1$ by Modified Look-Locker-Inversion Recovery techniques with balanced gradient echo readouts, i.e. classical MOLLI \cite{Messroghli2004} and its modifications, e.g. see Ref.~\cite{Piechnik2010}, as well as spoiled gradient echo readouts \cite{Gensler2015, Campbell2013}   has shifted into the focus of interest in cardiac MRI. The MOLLI-techniques with either readouts  differ from the aforementioned examples as the periodic perturbation acts on two time scales. Periods of free longitudinal relaxation, the length of which are determined by the heart beat cycle length $\TRR$, are interrupted by readout imaging modules, in which the balanced or spoiled gradient echoes are repeated with the much smaller repetition time $\TR$. Of course it would be of paramount interest to relate this complex driven relaxation process with its apparent relaxation time $T_1^*$ to the sequence parameters, and the tissue parameters $T_1$ and $T_2$. This dependence, which to our knowledge is still unknown, will be derived in this manuscript. 

\section{Longitudinal relaxation in the presence of discontinuous periodic readouts}
Periodic perturbation of relaxation processes consist of modules in which external forces, e.g. radio-frequency pulses, are interleaved with non-disturbed relaxation intervals in which thermodynamic forces act. The latter increase entropy of the spin system which becomes apparent in transverse relaxation, and minimize its free energy in longitudinal relaxation. Pulses act linearly on the magnetization vector $\m$, whereas thermodynamic forces on the difference of $\m$ to its equilibrium value 
$\m_\mathrm{eq} $, i.e.  $\m-\m_\mathrm{eq}$. For simplification we normalize the magnetization by the magnitude of this equilibrium value   ${\bf m}\to {\bf m}/m_\mathrm{eq}$, and align the $z$-direction parallel to the direction of the external magnetic field, i.e.  $\m_\mathrm{eq}\to \ez$, with $\ez$ as the corresponding unit vector. 

The above mentioned linear/affine response of the magnetization to rf-pulses and thermodynamic forces has the following consequence: when rf-puls(es) and relaxation period are coupled to one module, and when these modules appear contiguously in series, the magnetization at the end of one module is an affine function of that at its beginning, i.e. that at the end of the preceding module. So, for the magnetization after the $n$th module follows
\begin{equation}
\m_{n}=\U\m_{n-1}+\bf{v}\;,\label{recursive1}
\end{equation}    
with the transformation matrix $\U$, and some vector $\bf{v}$. The steady state is achieved, when the magnetization after the module is identical with that before, which determines the corresponding steady state magnetization as
\begin{equation}
\m_\mathrm{ss}=(\mathbf{1}-\U)^{-1}\bf{v}\;, \label{ss}
\end{equation}  
with $\mathbf{1}$ as the identity matrix. Recursive application of Eq.~(\ref{recursive1}) and applying rules for geometric series yields that 
\begin{equation}
\m_{n}=\U^{n}(\m(0)-\m_\mathrm{ss})+\m_\mathrm{ss}\;, \label{Ev1}
\end{equation} 
with $\m(0)$ as the initial magnetization. The last equation demonstrates that with respect to the steady state magnetization $\U$ is the generator of evolution on the time scale of a module duration.  For practical determination of relaxation rates,  $\U$ may be decomposed in its spectral components, after Eigenvectors and Eigenvalues have been determined. In general this yields a multi-exponential decay of $\m$.
  
In case of discontinuous Lock-Locker sequences like MOLLI the situation is a bit more complex as the module consists of two sub-modules, the readout, with duration $\tread$ and the free relaxation period, lasting $\tfree=\TRR-\tread$, with $\TRR$ as the cycle length of the heart beat. Within the module the sub-modules follow a time evolution as in Eq.~(\ref{Ev1}). So when $\m_{n-1}$ is the magnetization before the $n$th readout, it develops towards 
\begin{equation}
\m_{n-1/2} =\Uro(\m_{n-1}-\mro) +\mro\;,\label{ASSFP}
\end{equation}
with $\Uro$ as the transformation matrix -, and $\mro$ as the steady state magnetization of the readout. Note, that $n-1/2$ symbolizes the magnetization directly after the $n$th readout, but before the free evolution of the $n$th sequence cycle, completing only "half" the evolution. The explicit forms of the readouts have been determined in the past, e.g. see Refs.  \cite{Deichmann1992,Scheffler2003}, and will be used later on. Thereafter the magnetization freely decays, which is described by the relaxation rate matrix
\begin{equation}
\label{Ufr}
\Ufr =\begin{pmatrix}
e^{-\tfree/T_2 } & 0& 0\\
0 & e^{-\tfree/T_2} & 0 \\
0 & 0& e^{-\tfree/T_1}
\end{pmatrix}\ .
\end{equation}
So magnetization after the readout develops to
\begin{equation}
\m_{n}=\Ufr(\m_{n-1/2}-\ez)+\ez\;.\label{AFree}
\end{equation}
The Eqs.~(\ref{ASSFP},\ref{AFree}) yield the recursive dependence of magnetizations before and after the whole module as 
\begin{equation}
\m_{n}=\Ufr\Uro(\m_{n-1}-\m_\mathrm{ss}) +\m_\mathrm{ss}\;,\label{Evolutioncomb}
\end{equation}
with the steady state magnetization of the MOLLI sequence determined according to Eq.(\ref{ss})
\begin{eqnarray}
\m_\mathrm{ss}&\!\!\!=\!\!&\mro\!+\frac{1}{\mathbf{1}\!-\Ufr\Uro}(\mathbf{1}\!-\Ufr)(\ez\!-\mro)\nonumber\\
&\!\!\!=\!\!&\frac{1}{\mathbf{1}\!-\Ufr\Uro}\bigg(\Ufr(1-\Uro)\mro\bigg.\nonumber\\
& & \big.\hspace{20ex}+(1-\Ufr)\ez\!\bigg)\label{ssMOLLI}\;.
\end{eqnarray}
This implies that the evolution operator on the time scale of the composite module $\TRR$ is 
\begin{equation}
\label{compositeG}
\boldsymbol{U}^{\hbox{\tiny {(comp)}}}=\Ufr\Uro\;.
\end{equation}
It is noteworthy that in case that the evolution operators $\Ufr$ and 
$\Uro$ commute, the relaxation rate(s) of the composite module are the time average of those of the sub-modules. This is easily seen as one can assign the evolution matrices $\boldsymbol{U}^{(i)}$  ($i=$ read -, free -, composite module) generator  matrices $\boldsymbol{R}$ with $\boldsymbol{U}^{(i)}=\exp(\boldsymbol{R}^{(i)}t_i)$. As $\tread+\tfree=\TRR$ one obtains the addition theorem for respective generators 
\begin{equation}
\boldsymbol{R}^{\hbox{\tiny{(comp)}}}=\frac{\tread}{\TRR}\;\; \boldsymbol{R}^{\hbox{\tiny{(read)}}}+\frac{\tfree}{\TRR}\;\; \boldsymbol{R}^{\hbox{\tiny{(free)}}}\;.
\end{equation}
In the next sections the two different readout modules which are commonly used, the traditional Lock-Locker (FLASH) and the bSSFP readout, will be investigated.

\section{MOLLI with bSSFP readouts}
\subsection{Evolution under the influence of bSSFP}
In bSSFP, magnetization is excited by an initial preparation $\alpha/2$ pulse. Thereafter it develops gradient induced echoes which are all balanced and driven consecutively by alternating $\alpha$-pulses spaced by repetition time $\TR=2\TE$.  The theory of longitudinal relaxation under the influence of this sequence has been studied extensively in the past, e.g. \cite{Hargreaves2001,Scheffler2003,Schmitt2004}, and only essentials necessary for understanding of the paper are repeated here. The repeated application of pulses implies that the evolution operator $\UF$ consists of a sequence of identical operators giving the time evolution within the repetition time $\mathbf{\hat{A}}$, i.e. for $m$ repetitions 
\begin{equation}
\UF=\mathbf{\hat{A}}^m\;\mathbf{\hat{P}}_\mathrm{pre}\;,
\end{equation} 
where $\mathbf{\hat{P}}_\mathrm{pre}$ is the operator realizing the initial preparation, i.e. for an $\alpha/2$ pulse rotated around the $x$-axis
\begin{equation}
\mathbf{\hat{P}}_\mathrm{pre}=\begin{pmatrix}
1&0 &0\\
0 &\cos(\alpha/2)&\sin(\alpha/2)\\
0 & -\sin(\alpha/2) & \cos(\alpha/2)
\end{pmatrix}\;.\label{Pre}
\end{equation} 

The operator $\mathbf{\hat{A}}$ is built up from operators describing pulse related rotations $\alpha$, phase shifts of the rotation axis (due to phase cycles), precession (due to off-resonance) as well as free relaxation within $\TR$. We focus only on the on-resonant case, with rotations around the $x$-axis as well as alternating pulse directions, which derives $\mathbf{\hat{A}}$ as
\begin{equation}
\label{bSSFPA}
\mathbf{\hat{A}}=\begin{pmatrix}
-e^{-\frac{\TR}{T_2}}&0 &0\\
0 &e^{-\frac{\TR}{T_2}}\cos(\alpha)&\!\!e^{-\frac{\TR}{2} \left(\frac{1}{T_1}+\frac{1}{T_2}\right)} \sin(\alpha)\!\!\\
0 & \!\!e^{-\frac{\TR}{2} \left(\frac{1}{T_1}+\frac{1}{T_2}\right)}\sin(\alpha)\!\! & e^{-\frac{\TR}{T_1}}\cos(\alpha)
\end{pmatrix}.
\end{equation} 
Note that this real matrix is symmetric (Hermitian), which implies an orthogonal system of Eigenvectors. The evolution of magnetization within $\TR$ is obtained by the 
 
\begin{equation}
{\bf m}_{m}^{\hbox{\tiny{(bSSFP)}}}=\mathbf{\hat{A}}\;{\bf m}_{m-1}^{\hbox{\tiny{(bSSFP)}}}+\bf{v}
\end{equation} 

 with 
 \begin{equation}
\bf{v}= \begin{pmatrix}
0\\
e^{-\frac{\TR}{2 T_2}}\left(1-e^{-\frac{\TR}{2 T_1}}\right)\sin(\alpha)\\
1+e^{-\frac{\TR}{2 T_1}}(\cos(\alpha)-1)-e^{-\frac{\TR}{ T_1}}\cos(\alpha)
\end{pmatrix}\ .
\end{equation}  
For the evaluation of longitudinal relaxation in the MOLLI setup, it is useful to obtain the steady state vector, as well as Eigenvectors and - values of the bSSFP readout matrix $\mathbf{\hat{A}}$. as the repetition time is much smaller than the relaxation time, $\TR\ll T_1,\;T_2$, we get
\begin{alignat}{3}
         \mathbf{e}_1 &=\begin{pmatrix}
                                1\\0\\0
                      \end{pmatrix}        & \quad & \quad &  \lambda_1 &=-\exp\left(-\frac{\TR}{T_2}\right)\nonumber\\
         \label{Eigenall}
         \mathbf{e}_2 &=\begin{pmatrix} 
                                0\\ -\cos(\alpha/2)\\ \sin(\alpha/2)
                      \end{pmatrix}        & & &  \lambda_2 &=-\exp\left(\frac{-\TR}{\Tzsb}\right)\\
         \mathbf{e}_3 &=\begin{pmatrix} 
                                0\\ \sin(\alpha/2)\\ \cos(\alpha/2)
                      \end{pmatrix}        & & &    \lambda_3 &=\exp\left(\frac{-\TR}{\Tsb}\right)\nonumber
\end{alignat}
for the normalized orthogonal Eigenvectors (left), and corresponding Eigenvalues (right). Here 
\begin{eqnarray}
\frac{1}{\Tsb}&=&\cos^2(\alpha/2)\frac{1}{T_1}+\sin^2(\alpha/2)\frac{1}{T_2}\cr
\frac{1}{\Tzsb}&=&\cos^2(\alpha/2)\frac{1}{T_2}+\sin^2(\alpha/2)\frac{1}{T_1}\label{bSSFPT}
\end{eqnarray}
denote the apparent longitudinal or transverse relaxation times of the bSSFP train with respect to the direction of the steady state magnetization, as the latter is parallel to the 3rd Eigenvector $\mathbf{e}_3$ and takes the form 
\begin{equation}
\mF\approx\cos(\alpha/2)\;\frac{\Tsb}{T_1 }\mathbf{e}_3\;. \label{ssbSSFP}
\end{equation}

\subsection{Evaluation under the influence of discontinuous bSSFP readouts}
We will now investigate the generator of time evolution in the MOLLI setup.
The generator for the bSSFP imaging module is given by
\begin{align}
    \Uro = \mathbf{\hat{P}}_\mathrm{post}\;\UF\;, 
\end{align}
where $\mathbf{\hat{P}}_\mathrm{post}$ denotes an operator which describes how magnetization is treated at the end of an imaging module. If the magnetization just remains, $\mathbf{\hat{P}}_\mathrm{post}$ is just the identity matrix. If magnetization is flipped back  onto the $z$-axis, $\mathbf{\hat{P}}_\mathrm{post}$ becomes a rotational matrix.  
Hence, the generator of the composite module, i.e. readout module followed by free relaxation is (Eq.~\eqref{compositeG}) 
\begin{equation}
\boldsymbol{U}^{\hbox{\tiny {(comp)}}}=\Ufr\Uro=\Ufr\;\mathbf{\hat{P}}_\mathrm{post}\;\mathbf{\hat{A}}^m\;\mathbf{\hat{P}}_\mathrm{pre}\;.\label{EvMOLLI}
\end{equation}
 
The free relaxation following the bSSFP readout lasts long, when compared to transverse relaxation $\TRR-\tread\gg T_2$. According to Eq.~\eqref{Ufr} this simplifies the corresponding evolution operator to 
\begin{equation}
\Ufr\approx e^{-(\TRR-\tread)/T_1}\;\mathbf{\hat{\Pi}}_z \;,\label{Ufrs}
\end{equation} 
where, with $\ez$ as unit vectors in $z$-direction   
\begin{equation}
\mathbf{\hat{\Pi}}_z=\ez\ez^{\bf{T}}=\begin{pmatrix}
0&0 &0\\
0 &0&0\\
0 & 0 & 1
\end{pmatrix}\;
\end{equation}
is the projection operator of a vector onto the $z$-axis. Note that $\ez^{\bf T}$ is the transposed vector. Keeping in mind that the initial magnetization is parallel to the $z$-axis, and that the free relaxation periods just leave a magnetization in $z$-direction (Eq.~\eqref{Ufrs}), it is sufficient to consider of $\mathbf{\hat{P}}_\mathrm{pre}$ in Eq.~(\ref{Pre}) only that part, which rotates the $z$-component of magnetization. So, together with the Eigenvectors in Eqs.~(\ref{Eigenall}) we may write
\begin{equation}
\mathbf{\hat{P}}_\mathrm{pre}=\e3\ez^{\bf T}\, .
\end{equation}   

Exploiting now that $\mathbf{e_3}$ is the 3rd Eigenvector of the bSSFP readout matrix $\mathbf{\hat{A}}$ (see Eqs.~\eqref{bSSFPA}, \eqref{Eigenall}), Equation~\eqref{EvMOLLI} is simplified to
 \begin{equation}
 \label{EvMolli2}
 \Ufr\UF=\mathbf{\hat{\Pi}}_z\ \xi\;\lambda_3^m\; e^{-\frac{\TRR-\tread}{T_1}}\;,
 \end{equation}
with 
\begin{equation}
\xi=\ez^{\bf T}\mathbf{\hat{P}}_\mathrm{post} \e3\, .
\end{equation}
As already mentioned above two options are considered for the fate of the magnetization immediately after the readout module. When there is no further manipulation we have simply $\mathbf{\hat{P}}_\mathrm{post} = \mathbf{1}$, with $\mathbf{1}$ as the identity matrix. If the magnetization is flipped back, the bSSFP Eigenvector $\e3$ is rotated back parallel to the $z$-axis, i.e. one gets $\mathbf{\hat{P}}_\mathrm{post}\e3 = \ez$. So 
\begin{equation}
\xi=\begin{cases}
\cos(\alpha/2) &\text{no flip back}\\
1 &\text{with flip back}\, .
\end{cases}
\end{equation}
 
With the apparent relaxation time  $\TsM$, the relaxation of the $z$-component within the duration of the composite module, i.e.  $\TRR$, is given by the factor $e^{-\TRR/\TsM}$. So, after inserting the 3rd Eigenvalue $\lambda_3$ of Eqs.~\eqref{Eigenall}) into Eq.~\eqref{EvMolli2}, and taking into account that $\tread=m\TR$ one obtains
\begin{equation}
\label{MOLLIT1Stern}
\frac{1}{\TsM}=\underbrace{\frac{T_\mathrm{RR} - \tread}{T_\mathrm{RR}}\frac{1}{T_1}+\frac{\tread}{T_\mathrm{RR}}\frac{1}{\Tsb}}_{\hbox{\tiny{= time averaged relaxation rate}}}- \frac{\ln(\xi)}{T_\mathrm{RR}}\;. 
\end{equation}
So the apparent relaxation rate is the sum of the time averaged rate of relaxation within the readout module and free relaxation plus a correction term, which depends on the post preparation. With Eq.~(\ref{bSSFPT})this leads to 

\begin{equation}
\label{MOLLIT1Stern2}
\frac{1}{\TsM}=\frac{1}{T_1}+\frac{\tread}{T_\mathrm{RR}}\sin^2\left(\alpha/2\right)\left(\frac{1}{T_2}-\frac{1}{T_1}\right)-\frac{\ln\left(\xi\right)}{T_\mathrm{RR}}\ .
\end{equation}

The correction term $\ln\left(\xi\right)/\TRR$ either vanishes if the flip back  is applied or is otherwise also rather small as the following shows: $\TRR$ is typically in the order of $T_1$ and bSSFP readout mainly operates with angles of $\alpha\approx 35^\circ$, which makes the relative difference of relaxation rates with and without flip back about $\approx 5\%$.

The steady state magnetization derives from Eq.~(\ref{ssMOLLI}). As at this time point only the $z$-component is important, one may simplify this Equation and derives 
\begin{eqnarray}
\label{Mss1}
m_{\mathrm{ss}}^{\hbox{\tiny{(MOLLI)}}}	&\!\!\!\!=\!\!\!& \frac{1}{1-e^{-\frac{\TRR}{\TsM}}}\Bigg(1-e^{-\frac{\TRR-\tread}{T_1}}\Bigg.\nonumber\\
& & \hspace{5ex}\Bigg.+ e^{-\frac{\TRR-\tread}{T_1}}\!\bigg(\!1\!-e^{\!-\frac{\tread}{\Tsb}}\!\bigg)\xi m_{\mathrm{ss}}^{\hbox{\tiny{(bSSFP)}}}\!\!\Bigg) \nonumber\\[1.5ex]
									 	&\!\!\!\!=\!\!\!&\frac{1\!-e^{\!\!-\frac{\TRR-\tread}{T_1}}\!\!\left(\!1\!-\xi m_{\mathrm{ss}}^{\hbox{\tiny{(bSSFP)}}}\!\! \left(\!1\!-e^{\!\!-\frac{\tread}{\Tsb}}\right)\!\!\!\right)}{1-e^{\!\!-\frac{\TRR}{\TsM}}}.\nonumber\\[-2ex]
\end{eqnarray}

When we assume that $\tread/\Tsb$ is sufficiently small, and inserting the steady state magnetization under bSSFP readout conditions $m_\mathrm{ss}^{\hbox{\tiny{(bSSFP)}}}$ from Eq.~(\ref{ssbSSFP}) one obtains
in first order expansion in $\tread$
plus a correction term, which depends on the post preparation. With Eq.~(\ref{bSSFPT})this leads to 

\begin{align}
m_\mathrm{ss}^{\hbox{\tiny{(MOLLI)}}}&\approx \frac{1-e^{-(\TRR-\tread\kappa\sin^2(\alpha/2))/T_1}}{1-e^{-\TRR/\TsM}}
\end{align}
with $\kappa = 1$ in the absence, and $\kappa = 1/2$ in the presence of the flip back of magnetization after the readout.
Complete neglection of the $\tread$ terms further simplifies these results to 
\begin{eqnarray}
\label{approxmss}
    m_\mathrm{ss}^{\hbox{\tiny{(MOLLI)}}}&\approx&\frac{1-e^{-\TRR/T_1}}{1-e^{-\TRR/\TsM}}
\end{eqnarray}
which has the similar form as the classical Lock Locker (FLASH) experiment hence motivating the commonly used correction
to obtain $T_1$ from the apparent relaxation time $\TsM$.

Equation (\ref{Mss1}) gives the steady state magnetization at the end of the free relaxation period, i.e. immediately before the next readout module. However, from this next readout module the relevant signal is that obtained when the center of k-space is acquired, the timing $\tim$ of which is (due to the preparation pulses) after half the duration of the readout module. Hence, one must also consider the effect of this bSSFP module and one obtains for the measured steady state magnetization    
\begin{align}
\label{msstim}
{\bf m}_{\mathrm{ss},\hbox{\tiny{im}}}^{\hbox{\tiny{(MOLLI)}}} &\!\!= {\bf m}_\mathrm{ss}^{\hbox{\tiny{(bSSFP)}}}+ (m_\mathrm{ss}^{\hbox{\tiny{(MOLLI)}}}\e3 - \m_\mathrm{ss}^{\hbox{\tiny{(bSSFP)}}})e^{-\frac{\tim}{\Tsb}}\cr\cr
                          & \!\!\approx \e3\left(\cos\left(\frac{\alpha}{2}\right)\frac{\tim}{T_1}  +m_\mathrm{ss}^{\hbox{\tiny{(MOLLI)}}}e^{-\frac{\tim}{\Tsb}}\right).
\end{align}
Note, that $\Tsb$ may be obtained from Eq.~\eqref{MOLLIT1Stern}.

\section{MOLLI with spoiled gradient echo readouts - discontinuous classical Look-Locker (FLASH)}
The theory of longitudinal relaxation under the influence of continously applied spoiled gradient echoes has been studied extensively in the past, e.g. \cite{Deichmann1992}. The spoiling implies that ideally the hf-pulses solely act on a magnetization in $z$-direction. So, within the readout module it is sufficient for two consecutive pulses to evaluate solely the interdependence of their $z$-component. As, in addition, the free relaxation period in between the readout module also only leaves a $z$-component at its end, it is justified as well to study only the $z$-component of the composed process.  This simplifies the mathematical analysis as matrix operations just reduce to multiplication with numbers.

\subsection{Evolution under the influence of spoiled gradient echos}
We will only roughly present the well known results.
The $z$-magnetization before two pulses in sequence, separated by the repetition time $T_\mathrm{R}$, are interrelated by the affine recursion
\begin{equation}
m_{m}^{\hbox{\tiny{(LL)}}}=A\;m_{m-1}^{\hbox{\tiny{(LL)}}}+v
\end{equation} 
 with 
 \begin{equation}
 A=e^{-\frac{T_\mathrm{R}}{T_1}}\cos(\alpha)
 \end{equation}
and
 \begin{equation}
v= 1-e^{-\frac{T_\mathrm{R}}{T_1}}\;.
\end{equation}
Note that the equilibrium magnetization is normalized to one. 
Recursive application directly leads to an apparent relaxation rate
\begin{eqnarray}
\label{T1SternLL}
\frac{1}{\TsLL}&=&\frac{1}{T_1}-\frac{\ln\left(\cos\left(\alpha\right)\right)}{T_\mathrm{R}}\;,\end{eqnarray}
and the steady state magnetization derives as
\begin{align}
\label{mssll}
    m_{\mathrm{ss}}^{{\hbox{\tiny (LL)}}} = \frac{1-e^{-T_\mathrm{R}/T_1}}{1-e^{-T_\mathrm{R}/\TsLL}}\approx  \frac{\TsLL}{T_1}\;.
\end{align}
The last approximation is justified, as the repetition time of the gradient echoes is very small compared with the native relaxation time $T_1$ and its apparent partner $T_1^{*}$. The latter holds according to Eq.~\eqref{T1SternLL} as flip angles $\alpha$ used in spoiled gradient echo imaging are normally very small. 

\subsection{Evaluation under the influence of discontinuous spoiled gradient echo readouts}
Time evolution during the free relaxation periods between the readouts is given by the factor  
\begin{equation}
U^{\hbox{\tiny{(free)}}}= e^{-(\TRR-\tread)/T_1}\label{Ufrs2}\;,
\end{equation}
where duration of the readout $\tread$ is determined by the number $m$ of gradient echoes, i.e. $\tread=m\TR$.
So, the generator of discontinuous relaxation (Eq.~(\ref{Evolutioncomb})) for one period (readout- free relaxation, with duration $\TRR$) in the Lock-Locker setup is 
\begin{eqnarray}
U^{\hbox{\tiny{(free)}}}U^{\hbox{\tiny{(LL)}}}&=&U^{\hbox{\tiny{(free)}}}\;A^m\label{EvLL}\nonumber\cr\cr
&=&e^{-(\TRR-\tread)/T_1}\;e^{-\tread/\TsLL}\nonumber\cr\cr
&=&e^{-T_\mathrm{RR}/\TsNCLL}
\end{eqnarray}
with the apparent relaxation rate of the discontinuous spoiled gradient echo readouts
\begin{align}
\frac{1}{\TsNCLL}&=\frac{T_\mathrm{RR} - \tread}{T_\mathrm{RR}}\;\frac{1}{T_1}+\frac{\tread}{T_\mathrm{RR}}\;\frac{1}{\TsLL}\label{timeavLL}\cr\cr
&=\frac{1}{T_1}-m\;\frac{\ln\left(\cos\left(\alpha\right)\right)}{T_\mathrm{RR}}\ .
\end{align}
Equation \eqref{timeavLL} implies that the apparent relaxation rate of the combined/discontinuous Look Locker process is the time average of the free relaxation rate and that of the readout module.    
The steady state magnetization derives from Eq.~(\ref{ssMOLLI}) as
\begin{eqnarray}
\label{MssNCLL}
m_\mathrm{ss}^{\hbox{\tiny{(NCLL)}}}&\!\!\!\!=\!\!\!& \frac{e^{\!-\frac{\TRR\!-\tread}{T_1}}\!\bigg(\!\!1\!-e^{\!-\frac{\tread}{\TsLL}}\! \bigg)m_\mathrm{ss}^{\hbox{\tiny{(LL)}}}\!\! +\!\bigg(\!\!1\!-e^{\!-\frac{\TRR\!-\tread}{T_1}}\!\!\bigg)}{1\!-e^{\!-\frac{\TRR}{\TsNCLL}}}\nonumber\\[1.5ex]
			&\!\!\!\!=\!\!\!&\frac{1-e^{\!-\frac{\TRR\!-\tread}{T_1}}\left(1- m_{\mathrm{ss}}^{\hbox{\tiny{(LL)}}} \left(1-e^{\!-\frac{\tread}{\TsLL}}\right)\!\!\right)}{1-e^{\!-\frac{\TRR}{\TsNCLL}}}\nonumber\\[-2ex]
\end{eqnarray}
and when we assume that $\tread/\TsLL$ is sufficiently small, one obtains
\begin{eqnarray}
m_\mathrm{ss}^{\hbox{\tiny{(NCLL)}}}&\approx&\frac{1-e^{-\TRR/T_1}}{1-e^{-\frac{\TRR}{\TsNCLL}}}\ .
\end{eqnarray}

As in the case of the bSSFP readout, one has to keep in mind that the above steady state magnetization is that at the end of the free relaxation period, just before the subsequent readout. This readout, or more precisely its timing of the center of k-space $\tim$ determines the measured steady state magnetization. Until the center of k-space is reached the steady state magnetization of Eq.~(\ref{ss}) evolves under the influence of repetitive spoiled gradient echos, i.e. measured steady state magnetization is obtained as  
\begin{align}
m_{\mathrm{ss,\hbox{\tiny{im}}}}^{\hbox{\tiny{(NCLL)}}} &= m_{\mathrm{ss}}^{\hbox{\tiny{LL}}} + \left(m_\mathrm{ss}^{\hbox{\tiny{(NCLL)}}}  - m_{\mathrm{ss}}^{\hbox{\tiny{(LL)}}}\right)e^{-\frac{\tim}{\TsLL}}\nonumber \\
\label{mssFLASHtim}
&\approx \frac{\tim}{T_1} + m_\mathrm{ss}^{\hbox{\tiny{(NCLL)}}}e^{-\frac{\tim}{\TsLL}}\;,
\end{align}
where Eq. \eqref{T1SternLL} can be used to eliminate $\TsLL$.

\begin{figure*}
\centering
\subfigure[]{\includegraphics[width=5.5cm]{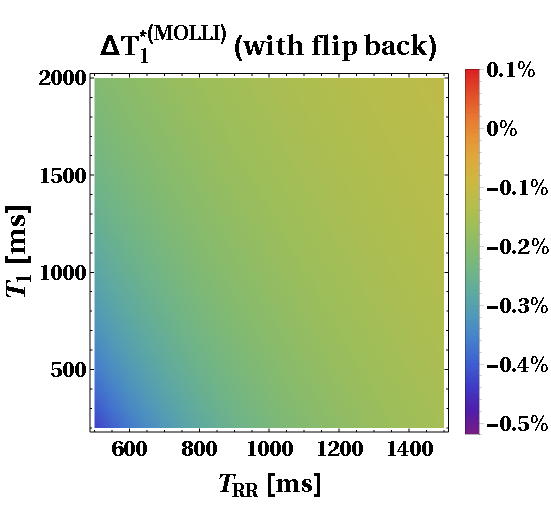}}
\subfigure[]{\includegraphics[width=5.5cm]{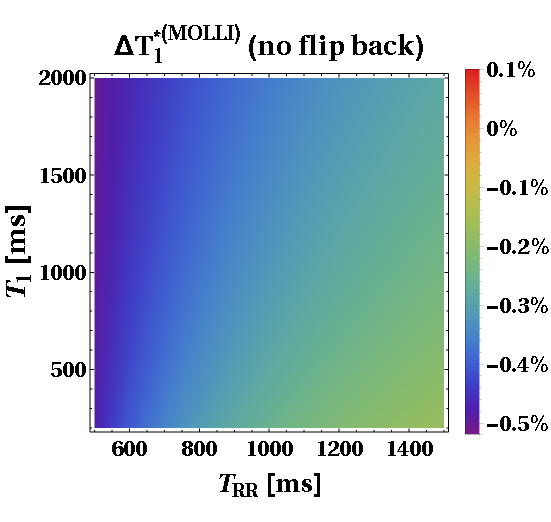}}
\subfigure[]{\includegraphics[width=5.5cm]{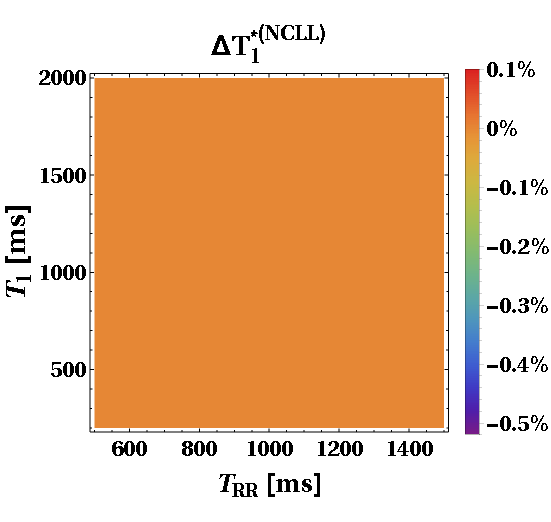}}\\
\subfigure[]{\includegraphics[width=5.5cm]{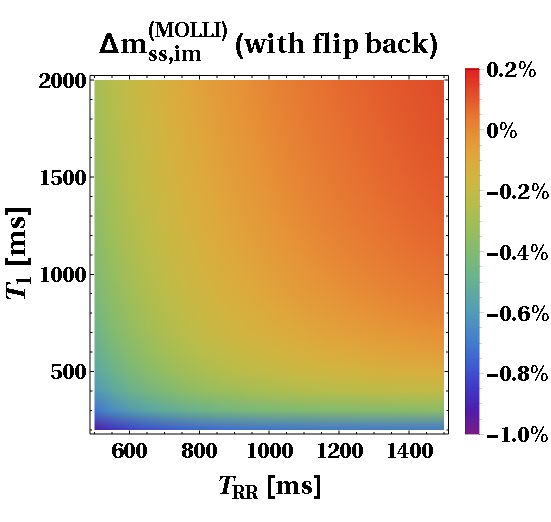}}
\subfigure[]{\includegraphics[width=5.5cm]{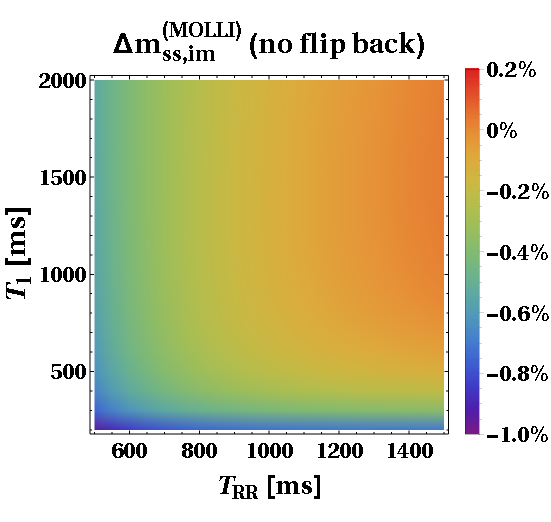}}
\subfigure[]{\includegraphics[width=5.5cm]{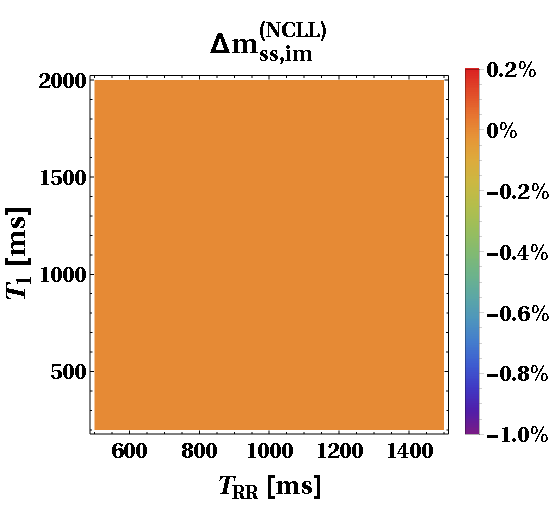}}
\caption{Relative error of analytical and numerical results for the MOLLI sequence with bSSFP readouts with and without flip back and the non continouse Look-Locker sequence with spoiled GRE readouts as a function of the $T_1$ and the heart cycle length $\TRR$. Above (a - c): for the apparent $\TsM$ and $\TsNCLL$ (see Eqs.~\eqref{MOLLIT1Stern},\eqref{MOLLIT1Stern2},\eqref{timeavLL} for the analytical values). Below (d - f): for the steady state magnetization (see Eqs.~\eqref{msstim},\eqref{mssFLASHtim}). For the numerical approach we straightforwardly applied the evolution matrices (e.g. preparation pulse, bSSFP/spoiled GRE readouts and free relaxation) on magnetization in series. Acquisition of the center of k-space was obtained at $\tread/2$, which was also the value for the imaging time $\tim$ determining the steady state magnetization (Eqs.~\eqref{msstim},\eqref{mssFLASHtim}). The time course of signal from these centers of k-space was fitted by a single exponential providing the numerical values for the apparent relaxation time and steady state magnetization. The MOLLI sequence parameters were: $\TR =$ 2.4 ms, $\alpha = 35^\circ$, $\tread =86 \text{\ pulses}\times 2.4 \text{\ ms} = \text{206.4 ms}$, and $T_2=50$ ms, and that for the spoiled GRE: $\TR =$ 2.4 ms, $\alpha = 7^\circ$, $\tread =80 \text{\ pulses}\times 2.4 \text{\ ms} = \text{206.4 ms}$, and $T_2=50$ ms. Note that the error for the GRE readout almost vanishes.}  
\label{testT1MOLLIwithFB}
\end{figure*}

\section{Summary and discussion}
We derived analytical expressions for the determination of the apparent longitudinal relaxation time in the presence of discontinuous bSSFP or spoiled gradient echo readouts. It turns out that the corresponding relaxation rates are approximately the time average of the rates during the readouts, for which expressions exist, and the free relaxation period (see Eqs. \eqref{MOLLIT1Stern} and \eqref{timeavLL}). Figure 1 demonstrates that the analytical results are close to those obtained by numerical simulations. Minor deviations may be due to the fact, that in case of the bSSFP readout the expressions for the transverse and longitudinal relaxation rates in direction of the steady state magnetization (Eqs.~\eqref{bSSFPT}) are already approximations which were obtained under the assumption $T_R\ll T_1,\;T_2$. 
In contrast, no such assumptions were made for the spoiled GRE readout, which explains that there is no difference besides numerical accuracy between numerical and analytical results. 
Our results now allow to determine the real $T_1$ from its apparent measured partner and sequence parameters, if the heart rate is known.

Our results also help to evaluate the existing techniques which assess $T_1$. The standard bSSFP MOLLI approaches rely on a 3-parameter fit $m(t)=A-B\;\exp(-t/\TsM)$, with the fit parameters  - apparent relaxation time $\TsM$,  $A$ as the steady state magnetization, which in our ansatz would correspond to $A=m_{\mathrm{ss},\hbox{\tiny{im}}}^{\hbox{\tiny{(MOLLI)}}}$ in Eq.~\eqref{msstim}, and $B$ as the dynamic range, i.e. the sum of equilibrium -- and steady state magnetization, $B=1+m_{\mathrm{ss},\hbox{\tiny{im}}}^{\hbox{\tiny{(MOLLI)}}}$ (note that we normalized the equilibrium magnetization to 1). The real $T_1$ is then  obtained by $T_1=(B/A-1)\TsM$ \cite{Messroghli2004}, i.e. 
\begin{equation}
\label{Messroghli}
T_1=\frac{\TsM}{m_{\mathrm{ss},\hbox{\tiny{im}}}^{\hbox{\tiny{(MOLLI)}}}}\Bigg.\Bigg|_{\hbox{fitted}}\, .
\end{equation} 
This formula is inferred from results valid in the setting of continuous application of spoiled gradient echoes. Here, in the limiting case for repetition times much shorter than the apparent $T_1$, a similar relationship between apparent and real $T_1$, as well as steady state magnetization holds (see Eq.~\eqref{mssll}). However, it is never questioned whether the prerequisites for this approximation are fulfilled. As the standard bSSFP MOLLI uses discontinuous readouts, it treats these readout modules as ``super'' pulses and the heart cycle length as repetition time, i.e, $(\TR)_{\hbox{standard approch}}\to \TRR$. However, as the heart cycle length and apparent $T_1$ are in the same order of magnitude, the necessary criteria, namely that $\TRR/\TsM\ll 1 $, is not fulfilled. Instead our ansatz yields from Eq.~\eqref{approxmss}, that steady state magnetization does not fulfill the above assumed Eq.~\eqref{Messroghli} but the inequility $m_\mathrm{ss}^{\hbox{\tiny{(MOLLI)}}}> \TsM/T_1$, i.e. true putative $T_1$ values obtained from Eq.~\eqref{Messroghli} may significantly differ from their true value. 

The question is, why does the standard MOLLI evaluation despite these obvious wrong presuppositions yield rather acceptable $T_1$ values?  The answer lies in the time $\tim$ at which the center of k-space is acquired. According to Eq.~\eqref{msstim} it locates the measured steady state magnetization $m_{\mathrm{ss},\hbox{\tiny{im}}}^{\hbox{\tiny{(MOLLI)}}}$ somewhere beneath that after the free relaxation period ($m_\mathrm{ss}^{\hbox{\tiny{(MOLLI)}}}$ in Eq.~\eqref{approxmss}), and above the steady state magnetization of the continuous bSSFP sequence ($m_\mathrm{ss}^{\hbox{\tiny{(bSSFP)}}}$ in Eq.~\eqref{ssbSSFP}), i.e. closer to the assumed $\TsM/T_1$. 
In fact, when we assume that the center of k-space is acquired at half the duration of the readout module, i.e. at $\tim=\tread/2$, and when we further take into account that $\tread\ll T_1$, which is in general a rather generous concession, the first term of the approximation of Eq.~\eqref{msstim}, $\cos(\alpha/2)\; \tim/T_1$, may be neglected. We then can write 
\begin{align}
m_{\mathrm{ss},\hbox{\tiny{im}}}^{\hbox{\tiny{(MOLLI)}}}&\approx m_\mathrm{ss}^{\hbox{\tiny{(MOLLI)}}}\exp\left(-\frac{1}{2}\;\frac{\tread\ \ }{\Tsb}\right)\cr
&\approx \frac{\sinh\left(\frac{1}{2}\;\frac{\TRR}{T_1}\right)}{\sinh\left(\frac{1}{2}\;\frac{\TRR\;\;}{\TsM}\right)}\;,
\end{align} 
where we inserted $m_\mathrm{ss}^{\hbox{\tiny{(MOLLI)}}}$ from Eq.~\eqref{Mss1} and $\Tsb$ from Eq.~\eqref{MOLLIT1Stern}, and made again use of $\tread\ll T_1$. With the weaker presuppositions, $1/2\;\TRR\ll T_1,\TsM$,  instead of $\TRR\ll T_1,\TsM$, it is at least understandable that expansion of the hyperbolic sinus provides the approximation of Messrhogli et al. , i.e. 
\begin{equation}
m_{\mathrm{ss},\hbox{\tiny{im}}}^{\hbox{\tiny{(MOLLI)}}}\approx \frac{\TsM}{T_1}\;.
\end{equation}
So the acceptable quality of the Eq.~(\ref{Messroghli}) for determination of $T_1$ results rather from serendipity than from  a rigorous based foundation. One might question, what would happen, if the center of k-space is acquired considerably prior to $\tread/2$.

In our opinion our results will be helpful for analysis of already existing $T_1$ mapping techniques as well as for the design of new ones. Also numerical approaches and simulation may be validated with these rather simple analytical expressions.

\subsection*{Acknowledgments}
The authors' research was supported by the Deutsche Forschungsgemeinschaft (SFB 688, TP B05 to WB) and the Bundesministerium f\"ur Bildung und Forschung (BMBF 01 EO1504 to WB).

\section*{References}

\bibliography{mybibfile}

\end{document}